\documentclass[aps,prl,twocolumn,showpacs]{revtex4-1}
\usepackage{amssymb}
\usepackage{graphicx}
\usepackage{dcolumn}
\usepackage{bm}
\usepackage{amsmath}
\usepackage{soul,color}
\usepackage{textcomp}
\usepackage{times}

\definecolor{cGreen}{RGB}{0,0,0}
\definecolor{cBlue}{RGB}{45,51,180}
\definecolor{cmagenta}{RGB}{205,0,100}

\usepackage[colorlinks,linkcolor=cBlue,citecolor=cBlue]{hyperref}

\begin{document}

\title{Synthetic U(1) Gauge Invariance in a Spin-1 Bose Gas}
\author{Chunping Gao$^{1}$}
\author{Jinghu Liu$^{1}$}
\author{Maolin Chang$^{1}$}
\author{Han Pu$^{2}$}
\email{hpu@rice.edu}
\author{Li Chen$^{1}$}
\email{lchen@sxu.edu.cn}

\affiliation{
$^1${Institute of Theoretical Physics and State Key Laboratory of Quantum Optics and Quantum Optics Devices, Shanxi University, Taiyuan 030006, China}\\
$^2${Department of Physics and Astronomy, and Rice Center for Quantum
    Materials, Rice University, Houston, TX 77005, USA}
}

\begin{abstract}
Recent experimental realizations of the lattice Schwinger model [Nature 587, 392 (2020) and Science 367, 1128 (2020)] open a door for quantum simulation of elementary particles and their interactions using ultracold atoms, in which the matter and gauge fields are constrained by a local U(1) gauge invariance known as the Gauss's law. Stimulated by such exciting progress, we propose a new scenario in simulating the lattice Schwinger model in a spin-1 Bose-Einstein condensate. It is shown that our model naturally contains an interaction of the matter fields which respects the U(1) gauge symmetry but has no counterpart in the conventional Schwinger model. In addition to the $\mathbb{Z}_2$-ordered phase identified in the previous work, this additional interaction leads to a new $\mathbb{Z}_3$-ordered phase. We map out a rich phase diagram and identify that the continuous phase transitions from the disordered to the $\mathbb{Z}_2$-ordered and the $\mathbb{Z}_3$-ordered phases belong to the Ising and the 3-state Potts universality classes, respectively. Furthermore, the two ordered phases each possess a set of quantum scars which give rise to anomalous quantum dynamics when quenched to a special point in the phase diagram. Our proposal provides a novel platform for extracting emergent physics in cold-atom-based quantum simulators with gauge symmetries.
\end{abstract}

\maketitle

\textit{Introduction ---}
Gauge invariance, which refers to the coordinated dynamics of matter and gauge fields being restricted by local symmetries at each spacetime location \cite{Weinberg1995}, has fundamentally shaped our understanding of interacting elementary particles in quantum electrodynamics (QED) \cite{Feynman2006} and quantum chromodynamics \cite{Marciano1978, Kogut1983, Ortmanns1996}. While a number of breakthroughs in synthesizing gauge fields in cold atoms have been made over the last decade \cite{Goldman2014}, including the experimental realization of artificial electric \cite{Lin2011-1} and magnetic fields \cite{Lin2009}, spin-orbit coupling \cite{Lin2011-2,Wang2012,Cheuk2012} and the density-dependent gauge field \cite{Gorg2019,Clark2018}, none of them is essentially endowed with local symmetry. Very recently, two experimental simulations of the lattice Schwinger model in cold atoms have changed the situation \cite{Yang2020,Mil2020}. In these experiments, the U(1) gauge symmetry is synthesized by locally tying the matter and gauge fields to each other via careful control of the tunneling and interactions of neural atoms. As a result, counterparts of such physical phenomena in particle physics as the spontaneous breaking of charge-parity symmetry \cite{Coleman1976}, string inversion and meson formation \cite{Banerjee2012, Pichler2016} are expected to be observed. Very recently, it has been shown that a Rydberg chain with nearest-site Rydberg blockade \cite{Bernien2017} can also be mapped to the U(1) lattice Schwinger model \cite{Surace2020}.

Motivated by the recent experimental progress, we propose a new platform to simulate the U(1) lattice Schwinger model in a spin-1 Bose-Einstein condensate (BEC). The significance of quantum simulator lies not only in simulating existing models of interest, but also in the emergent new physics arising from the intrinsic properties of the simulators. Here we show that the particle collisions in the spinor BEC naturally lead to a term corresponding to the matter-field interaction that has no counterpart in the conventional Schwinger model \cite{Yang2020,Mil2020,Coleman1976,Banerjee2012, Pichler2016}. Consequently, not only we recover the same phases (a disorder and a $\mathbb{Z}_2$ ordered phase) observed in previous simulators, but we also identify a new ordered phase breaking the $\mathbb{Z}_3$ translational symmetry. We prove that the second-order phase transitions from the disordered to the $\mathbb{Z}_2$-ordered and the $\mathbb{Z}_3$-ordered phases fall into the Ising and the 3-state Potts universality classes, respectively. The Potts criticality is intimately related to the anomalous quench dynamics and the quantum scars associated to the $\mathbb{Z}_3$-state, thus is of help in tracing the origin of $\mathbb{Z}_3$-related quantum scars. This new $\mathbb{Z}_3$-ordered phase exists in the experimentally realizable parameter regime of the commonly used atomic species such as $^{23}$Na and $^{87}$Rb \cite{Kawaguchi2012,Stamper-Kurn2013}, we therefore expect that these emergent physics can be experimentally observed in the near future.

\begin{figure}[t]
	\includegraphics[width=0.46\textwidth]{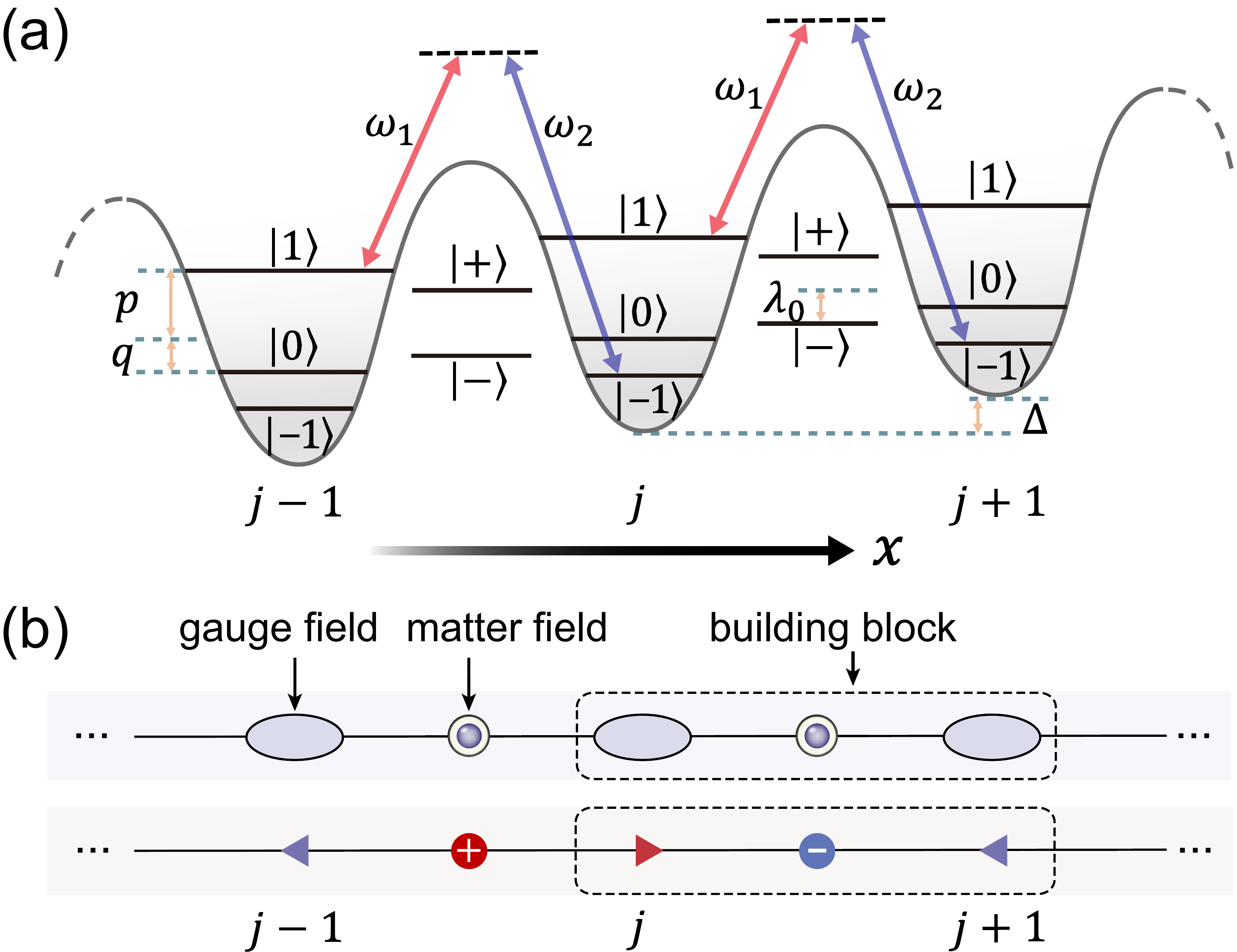}
	\caption{(a) Schematic of our model in the lab frame. $p$ and $q$ respectively denote the linear and quadratic Zeeman splittings, and $\Delta$ is a gradient potential. $\omega_1$ and $\omega_2$ are the frequencies of the lattice beam (on $x$ direction) and the traveling wave (on $z$ direction) which resonantly couple the states $\left\vert j+1,-1\right\rangle$ and $\left\vert j,1\right\rangle$ forming a $\Lambda$-type Raman process. (b) Upper panel: diagram of the U(1) lattice gauge model, where the gauge and the matter fields correspond to the bare state $\left\vert j,0\right\rangle$ and the dressed state $\left\vert j,-\right\rangle$ in subfigure (a), respectively. Lower panel: QED analog composed by electrons and positrons (matter fields), and electric fields (gauge fields). A building block surrounded by the dashed line consists of two neighboring gauge fields and one matter field.}
	\label{Fig1}
\end{figure}

\textit{Model ---}
We consider a spin-1 BEC deeply confined in a one-dimensional optical lattice along the $x$-direction, as is schematically shown in Fig.~\ref{Fig1}(a). Under the tight-binding approximation,  we label the lowest-band Wannier wave function of the site $j$ by $|j,\sigma\rangle$ with $\sigma=\{1,0,-1\}$ indicating the bare spin states. We construct the spin-dependent hopping using the technique of laser-assisted hopping \cite{Miyake2013}. To do so, we first introduce a biased magnetic field and a gradient potential, where the former defines a fixed quantization axis and provides the Zeeman shifts, and the latter provides a spin-independent tilt with strength $\Delta$. Next, we shine the BEC with a traveling light on the $z$-direction which, together with the lattice beam, forms a $\Lambda$-type Raman transition that couples the states $|j,\sigma=1\rangle$ and $|j+1,\sigma=-1\rangle$. Now, we write out the total Hamiltonian in the lab frame as (setting $\hbar=1$)
\begin{equation}
\hat{H} = \hat{H}_0 + \hat{H}_\text{int},
\label{H}
\end{equation}
with $\hat{H}_0$ the single-particle Hamiltonian given by
\begin{equation}
\begin{aligned}
\hat{H}_0 =&\sum_j \left[ p\hat{F}_j^z +q\hat{\Xi}_j +  j\Delta \hat{n}_j \right. \\
& \left. + (-1)^j\lambda_0 \left( e^{-i\delta\omega t} \hat{b}_{j,1}^\dagger \hat{b}_{j+1,-1} + \text {h.c.}\right) \right],
\end{aligned}
\label{H0}
\end{equation}
and $\hat{H}_\text{int}$ the interaction Hamiltonian \cite{Ho1998, Ohmi1998} given by
\begin{equation}
\hat{H}_\text{int}=\frac{U_{0}}{2}\sum_{j}\hat{n}_{j}\left( \hat{n}_{j}-1\right) +\frac{U_{2}}{2}\sum_{j}\left( \hat{\mathbf{F}}_{j}^{2}-2\hat{n}_{j}\right),
\label{Hint}
\end{equation}
where $\hat{F}_j^\mu \equiv \sum_{\sigma,\sigma^{\prime}} \hat{b}_{j,\sigma}^{\dagger}S_{\sigma\sigma^{\prime}}^\mu \hat{b}_{j,\sigma^{\prime}}$ are the local spin operators with $S^{\mu=x,y,z}$ the generalized spin-1 matrices, and $\hat{b}_{j,\sigma}$ is the bosonic field operator of spin $\sigma$ at site $j$, accordingly $\hat{n}_{j}=\sum_\sigma \hat{n}_{j,\sigma}$ is the local number operator with $\hat{n}_{j,\sigma} = \hat{b}_{j,\sigma}^\dagger \hat{b}_{j,\sigma}$, and $\hat{\Xi}_j = \hat{n}_{j,1} + \hat{n}_{j,-1}$. Here, $p$ and $q$ in $\hat{H}_0$ represent the linear and quadrtic Zeeman shift, respectively; $U_0$ and $U_2$ in $\hat{H}_\text{int}$ indicate the strengths of the spin-independent and the spin-dependent interaction, respectively. The last term in $\hat{H}_0$ characterizes the Raman coupling with $\delta\omega=\omega_2 - \omega_1$ the frequency difference between the two Raman beams, and $ (-1)^j\lambda_0$ the staggered hopping, where the phase factor $(-1)^j$ can be achieved by making the net recoil momentum acquired by the atoms equal one-half the lattice wave vector \cite{SM}.

We obtain the time-independent Hamiltonian in the rotating frame, i.e., $\hat{H}_0 \rightarrow \hat{U}(t) \hat{H}_0 \hat{U}^\dagger(t) - i \hat{U}(t)\partial_t \hat{U}^\dagger(t)$, where $\hat{U}(t)$ is a unitary operator properly chosen to eliminate both the phase factor $e^{i\delta\omega t}$ in the hopping term and the gradient term $\Delta$ \cite{SM}.
Furthermore, an additional transformation $\hat{b}_{j,1}\rightarrow e^{i(j\pi )}\hat{b}_{j,1}$ and $\hat{b}_{j,0}\rightarrow e^{i(j\pi/2)}\hat{b}_{j,0}$ is applied, which removes the stagger phase factor $(-1)^j$ in $\hat H_0$ while keeping $\hat H_\text{int}$ intact \cite{SM}. After these transformations, we have
\begin{equation}
\hat{H}_0 =\sum_j \left[p' \hat{F}_j^z +q\hat{\Xi}_j + \lambda_0 \left(\hat{b}_{j,1}^\dagger \hat{b}_{j+1,-1} + \text{h.c.}\right)\right],
\label{H0R}
\end{equation}
with $p'=p-\Delta/2-\delta\omega/2$. We will focus on the case under Raman resonance, i.e., $p'=0$. The resonant Raman coupling introduces two dressed states, denoted by $|\pm \rangle$ and gapped by $2\lambda_0$ as illustrated in Fig.~\ref{Fig1}(a), with the associated annihilation operators given by
\begin{equation}
    \hat{a}_{j,\pm} = \frac{1}{\sqrt{2}}(\hat{b}_{j,1}\pm\hat{b}_{j+1,-1})\,.
\end{equation}

Now we construct the U(1) lattice gauge model. First, we define the mode $\hat{b}_{j,0}$ on the spin-0 component as the \textbf{gauge field}, schematically denoted by the ovals in Fig.~\ref{Fig1}(b); and the lower-lying dressed mode $\hat{a}_{j,-}$ as the \textbf{matter field}, indicated by circles in Fig.~\ref{Fig1}(b). The spin-dependent interaction $U_2$ in Eq.~(\ref{Hint}) thus establishes the matter-gauge interaction. To prevent particles from being scattered into other modes (than the matter or the gauge modes defined above), we adopt the following two restrictions \cite{SM}: i) we require $\lambda_0\simeq q \gg U_0$, in which case two neighboring lower-energy dressed states $|- \rangle$ are resonantly coupled to the gauge field $|j,0 \rangle$ through the spin-exchange interaction, i.e., $\sim \hat{a}_{j-1,-}\hat{b}_{j,0}^\dagger \hat{b}_{j,0}^\dagger \hat{a}_{j,-}$, while the higher-energy dressed states $|+ \rangle$ are far off-resonant; ii) we restrict that there are at most two particles on a gauge mode and at most one particle on a matter mode. Restriction ii) can be satisfied by a proper preparation of the initial state. These two restrictions help to further simplify $\hat{H}$ and lead to the effective Hamiltonian  \cite{SM}
\begin{equation}
\begin{aligned}
\hat{H}_\text{eff} &=m\sum_{j}%
\hat{N}_{j}+\frac{\tilde{U}}{2}\sum_{j}\hat{N}_{j-1}\hat{N}_{j}\\
& -\frac{U_{2}}{2}\sum_{j}\left( %
\hat{a}_{j-1,-}\hat{b}_{j,0}^{\dagger } \hat{b}_{j,0}^{\dagger }\hat{a}_{j,-} +\text{h.c.}\right),
\end{aligned}
\label{heff}
\end{equation}
where we have defined $\hat{N}_{j} = \hat{a}_{j,-}^\dagger \hat{a}_{j,-}$ the number operator of the matter field, $m = q-\lambda_0-U_0/2$ and $\tilde{U} = (U_0-U_2)/2$.

One can observe that $\hat{H}_\text{eff}$ possesses a global translational symmetry and a local U(1) gauge symmetry, where the latter is generated by the Gauss operator $\hat{G}_j$:
\begin{equation}
\hat G_j=\hat{N}_{j}+\frac{\hat{n}_{j+1,0}+\hat{n}_{j,0}}{2}-1\,,
\label{G}
\end{equation}
which is defined on a building block consisting of two neighboring gauge fields and one matter field, as illustrated in Fig.~\ref{Fig1}(b). Furthermore, to acquire the QED interpretation of $\hat{H}_\text{eff}$, we perform the Jordan-Wigner transformation on the matter fields $\hat{a}_{j,-}$ and rewrite $\hat{H}_{\rm eff}$ as \cite{SM}
\begin{eqnarray}
\hat{H}_\text{f} &=& \left(\frac{\tilde{U}}{2}+m\right) \sum_{j} (-1)^j  \hat{\psi}_j^\dagger \hat{\psi}_j -  \frac{\tilde{U}}{2}\sum_{j} \hat{\psi}_{j-1}^\dagger \hat{\psi}_{j-1} \hat{\psi}_{j}^\dagger \hat{\psi}_{j} \notag \\
&-& \frac{U_2}{\sqrt{2}}\sum_{j} \left(\hat{\psi}_{j-1}^\dagger \hat \sigma_j^+ \hat{\psi}_{j}+ \text{h.c.}\right). \label{LSM}
\end{eqnarray}
Clearly, in the case of $\tilde{U}=0$, $\hat{H}_\text{f}$ reproduces the quantum link expression of the lattice Schwinger model with the gauge fields being realized by spin-1/2 spinors \cite{Yang2020,Wiese2013}. Specifically, the first term of $\hat H_\text{f}$ characterizes the staggered mass of the charged ferminons $\hat{\psi}_j$ and the last term denotes the matter-gauge interaction with $\hat{\sigma}^{\pm}_j$ the raising/lowering operators of the photons (gauge bosons) \cite{QLM}. $\hat{H}_\text{f}$ provides the following QED interpretation of $\hat{H}_\text{eff}$. The occupation of even and odd matter sites respectively represent the electrons and the positrons (see Fig.~\ref{Fig1}(b)), and the spin-exchange interaction ($U_2$ term) describes the process that a pair of electron and positron annihilate with each other and in the mean time the electric field is flipped. In a building block, the local Gauss operator $\hat{G}_j$ ensures the total flux of the electric field being equal to the number of charged particles, representing the manifestation of the Gauss's Law. For finite $\tilde{U}$, we additionally have a nearest-site matter-matter interaction (the second term in $\hat H_\text{f}$) that has no counterpart in the conventional Schwinger model \cite{Coleman1976}. This term comes from the intrinsic interactions of the spin-1 BEC and will lead to a rich phase diagram as will be shown below.

\begin{figure}[t]
\includegraphics[width=0.48\textwidth]{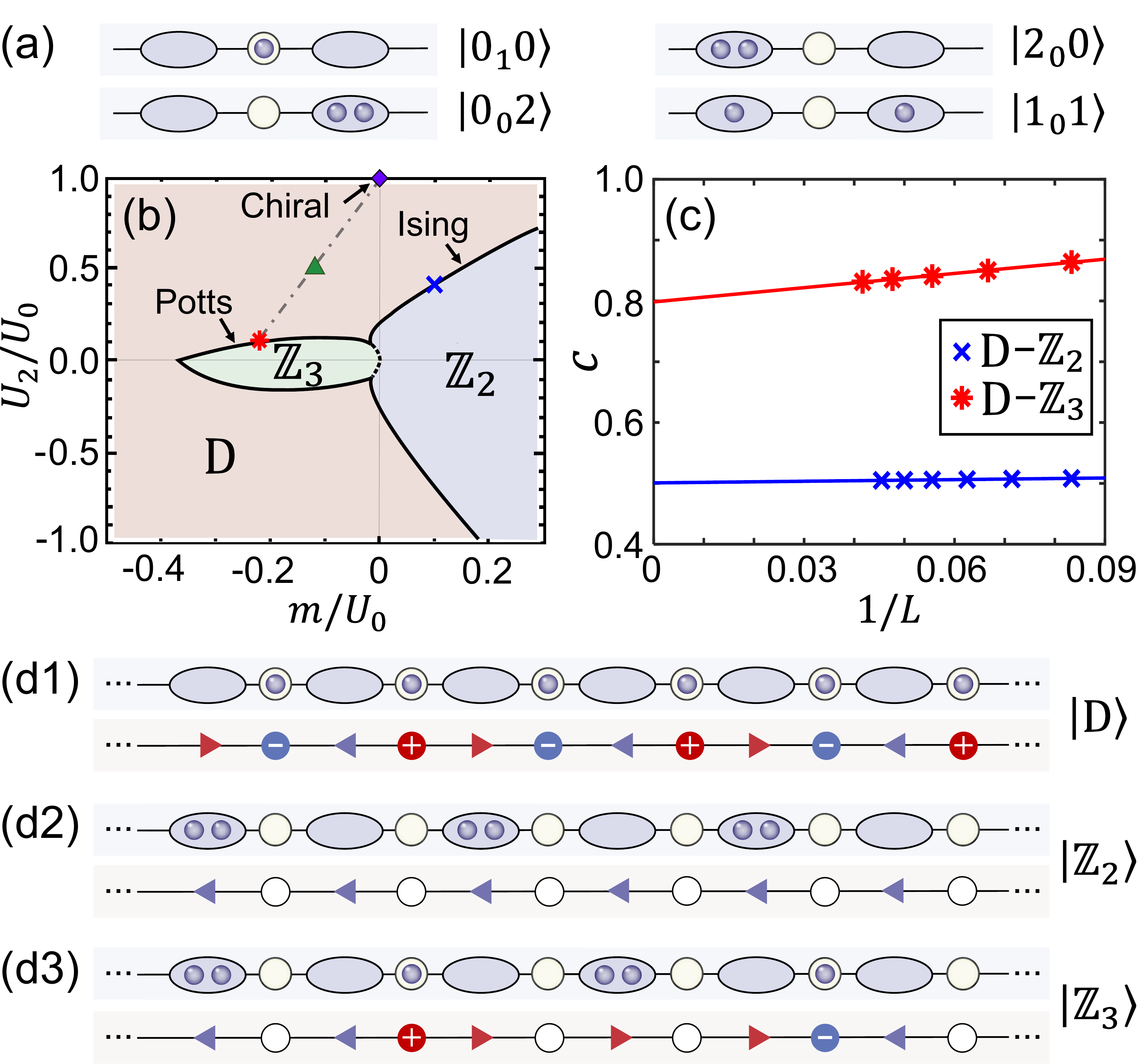}
\caption{(a) Four allowed configurations in a building block within the $G_j=0$ gauge sector. (b) Ground-state phase diagram in the $m$-$U_2$ plane, where the dashed and the solid lines respectively denote the first- and second-order phase boundary. In the diagram, the dot-dashed line satisfying $\tilde{U}=-2m$ connects the chiral-symmetric point (diamond) to the Potts critical point (star). On this line, we track the $|\mathbb{Z}_3\rangle$-related quantum scars as are discussed in Fig.~\ref{Fig3}. (c) Finite-size-scaling analysis of the central charge $c$ on the 2nd-order phase transitions D-$\mathbb{Z}_2$ and D-$\mathbb{Z}_3$ with the cross and the star corresponding to the critical points located at ($m\simeq0.1 U_0$, $U_2=0.4 U_0$) and ($m\simeq-0.22 U_0$, $U_2=0.1 U_0$) in (b), respectively. (d) Occupation configurations (upper line) and the corresponding QED analogy (lower line) of the three states $|D\rangle$, $|\mathbb{Z}_2\rangle$ and $|\mathbb{Z}_3\rangle$ at $U_2=0$. For $U_2 \neq 0$, population in individual sites are no longer conserved. However, the ground state degeneracy in each phase remains the same.}
\label{Fig2}
\end{figure}

\textit{Phase diagram---}
We discuss the equilibrium phases at 1/3 filling, i.e., there are totally $L$ particles for a chain with $L$ lattice sites, and focus on the gauge sector with no background charges, i.e., $G_j=0$ \cite{Coleman1976}. In this case, four occupation configurations, $\left|0 _1 0\right\rangle$, $\left|2 _0 0\right\rangle$, $\left|0 _0 2\right\rangle$ and $\left|1 _0 1\right\rangle$, are allowed in a building block, as displayed in Fig.~\ref{Fig2}(a), where $\left|n_{j,0} {}_{N_j} n_{j+1,0}\right\rangle$ denotes the Fock basis. Since the state $\left|1 _0 1\right\rangle$ is a dark state that are not coupled to the other three states through the $U_2$ interaction, we restrict our discussion within the subspace spanned by the remaining three states.

We plot the ground-state phase diagram in the $m$-$U_2$ plane in Fig.~\ref{Fig2}(b) obtained via numerically diagonalizing $\hat{H}_\text{eff}$ with $L=18$. A disordered phase D and two ordered phases $\mathbb{Z}_2$ and $\mathbb{Z}_3$ are identified. Three phases exhibit different ground-state degeneracy: the disordered phase D is non-degenerate, whereas the ordered phases $\mathbb{Z}_2$ and $\mathbb{Z}_3$ possess two- and three-fold degeneracy, respectively. In Figs.~\ref{Fig2}(d1)-(d3), we show particle number distributions of the three phases and their QED analog at $U_2=0$. Clearly, the phase D exhibits a configuration with all the matter fields being occupied whose wave function $|\text{D} \rangle=|\cdots 0_1 0_1 0_1 \cdots\rangle$ preserves the translational symmetry of $\hat{H}_\text{eff}$. On the other hand, the ordered phase $\mathbb{Z}_2$ ($\mathbb{Z}_3$) spontaneously breaks the translational symmetry in a $\mathbb{Z}_2$ ($\mathbb{Z}_3$) way such that the two (three) ground-state wave functions, $|\mathbb{Z}_2 \rangle = |\cdots 2_0 0_0 2_0 \cdots\rangle$ and $|\overline{\mathbb{Z}_2} \rangle= |\cdots 0_0 2_0 0_0 \cdots\rangle$ ($|\mathbb{Z}_3 \rangle = |\cdots 2_0 0_1 0_0 \cdots\rangle$, $|\overline{\mathbb{Z}_3} \rangle =|\cdots 0_0 2_0 0_1 \cdots\rangle$ and $|\overline{\overline{\mathbb{Z}_3}} \rangle =|\cdots 0_1 0_0 2_0 \cdots\rangle$), are energy-degenerate. We emphasize that in the conventional Schwinger model with $\tilde{U}=0$ (i.e., $U_0=U_2$), the $\mathbb{Z}_3$ phase is absent and only the D and the $\mathbb{Z}_2$ phases exist \cite{Yang2020,Coleman1976,Rico2014}. The occurrence of the $\mathbb{Z}_3$ phase results from the competition between the negative mass term $m<0$ and the repulsive matter-matter interaction ($\tilde{U}>0$) in $\hat{H}_{\rm eff}$: the former favors all the matter fields being occupied, while the latter hinders two neighboring matter fields being occupied simultaneously.

In the phase diagram, we identify phase transitions between the disordered phase and the ordered phases, D-$\mathbb{Z}_2$ and D-$\mathbb{Z}_3$, to be of 2nd order, while the transition between the two ordered phases $\mathbb{Z}_2$-$\mathbb{Z}_3$ of 1st order. The phase boundaries as well as the transition orders are determined by whether the 1st- or the 2nd-order derivatives of the ground-state energy with respect to the parameters ($m$ or $U_2$) exhibit discontinuity or not \cite{SM}. Furthermore, the 2nd-order transitions D-$\mathbb{Z}_2$ and D-$\mathbb{Z}_3$ respectively belong to the Ising and the 3-state Potts universality classes, whose low-energy critical behaviors are described by the conformal field theory with different central charges $c$ \cite{Francesco1997}. Practically, one can extract $c$ through fitting the curve \cite{Calabrese2004}
\begin{equation}
\mathcal{S}(l_A)=\frac{c}{3} \ln\left[\frac{L}{\pi}\sin\left(\frac{\pi l_A}{L}\right)\right]+s',
\label{CC}
\end{equation}
where $\mathcal{S}(l_A) =-\mathrm{Tr}( \hat{\rho}_A\log \hat{\rho}_A)$ is the von Neumann entropy of the subsystem A with length $l_A$, and $s'$ is a non-universal factor. In Fig.~\ref{Fig2}(c), we show the dependence of $c$ as a function of the chain length $L$ at two critical points (corresponding to the cross and the star in Fig.~\ref{Fig2}(b)), in which one can observe that the transitions D-$\mathbb{Z}_2$ and D-$\mathbb{Z}_3$ exhibit $c=0.5$ and $c=0.8$ in the thermodynamic limit $1/L\rightarrow 0$, clearly indicating the Ising and the Potts universality classes, respectively \cite{Francesco1997}. From experimental point of view, the intrinsic spin-dependent interaction for $^{23}$Na and $^{87}$Rb is $U_2/U_0 \approx 1\% $ and $ -0.5\%$ \cite{Kawaguchi2012,Stamper-Kurn2013}, respectively, and hence the emerged $\mathbb{Z}_3$ phase, D-$\mathbb{Z}_3$ and $\mathbb{Z}_2$-$\mathbb{Z}_3$ phase transitions are expected to be directly observed in these two most commonly used atomic species.

\begin{figure}[t]
\includegraphics[width=0.48\textwidth]{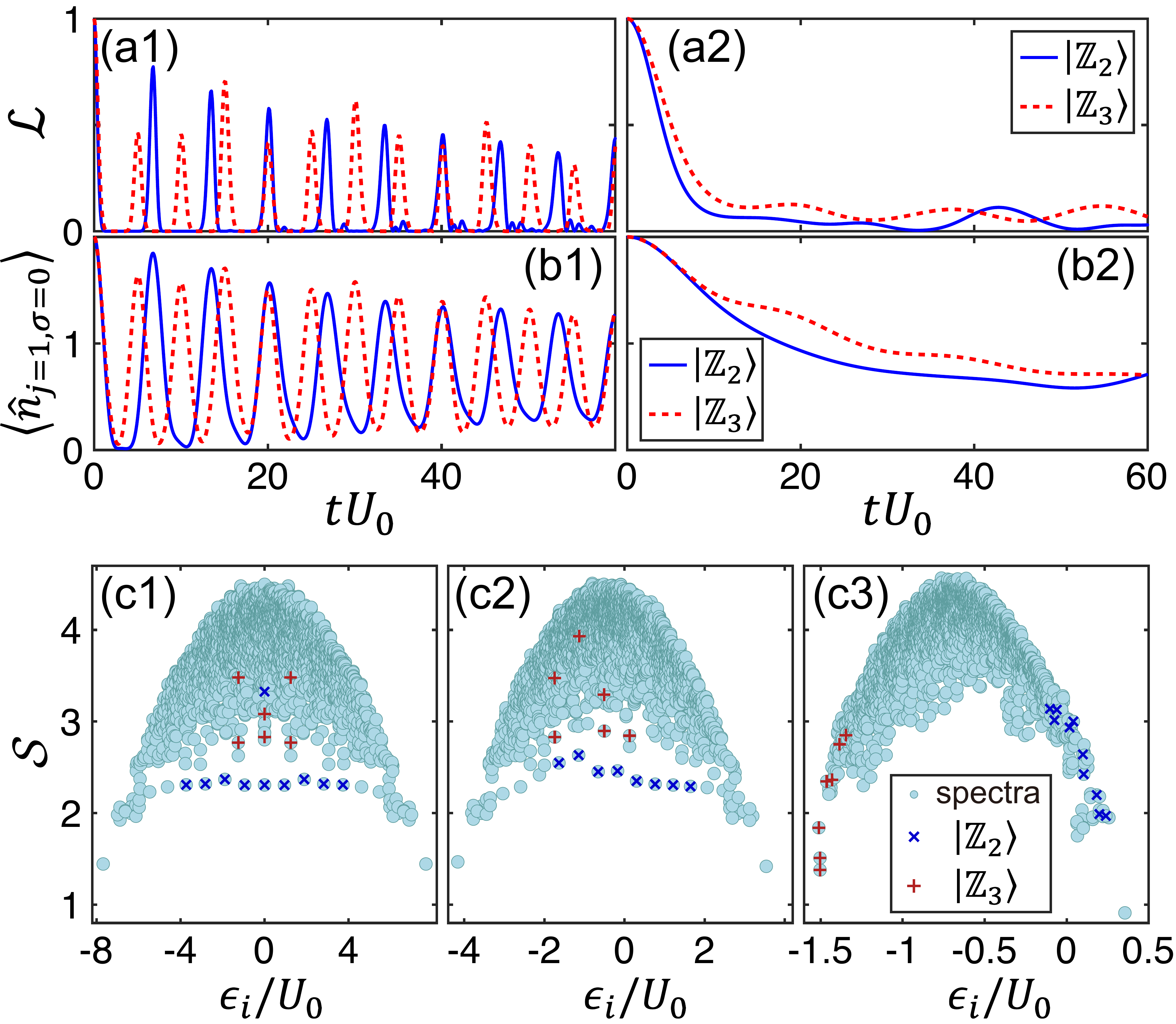}
\caption{Dynamics of the Loschmidt echo ${\cal L}$ (a) and occupation on the first gauge site (b) as $|\mathbb{Z}_2\rangle$ and $|\mathbb{Z}_3\rangle$ states are quenched to the chiral-symmetric point (a1, b1) and the Potts critical point (a2, b2). (c) The von Neumann entropy ${\cal S}$ plotted versus energy spectra $\epsilon_i$ at the chiral (denoted by diamond in Fig.~\ref{Fig2}(b)), the middle (triangle) and the Potts critical (star) points on the line of $\tilde{U}=-2m$ in (c1)-(c3), where crosses and plus signs indicate the scar states with large overlap to the $|\mathbb{Z}_2\rangle$ and $|\mathbb{Z}_3\rangle$ states, respectively. In the calculation, we take $L=18$ and ${\cal S}$ is calculated using $l_A=L/2$.}
\label{Fig3}
\end{figure}

\textit{Quench dynamics and quantum scars --- }Since we fix the gauge sector ($G_j=0$), the matter field and the gauge field are no longer independent. The U(1) lattice gauge model can therefore be mapped to a spin-1/2 chain by eliminating the matter fields \cite{Surace2020}, i.e., $
\hat{a}_{j-1,-}\hat{b}_{j,0}^{\dagger } \hat{b}_{j,0}^{\dagger }\hat{a}_{j,-} +\text{h.c.} \leftrightarrow \hat{\sigma}_j^x$ and $
\hat{N}_j \leftrightarrow -(\hat{\sigma}_j^z+\hat{\sigma}_{j+1}^z)/2$ using the Gauss's Law Eq.~(\ref{G}) with $\hat{n}_{j,0} \leftrightarrow 1+\hat{\sigma}_j^z$. Following this rule, the mapped spin Hamiltonian takes the form \cite{SM}
\begin{eqnarray}
\hat{H}_\text{s} &=& -m \sum_{j} \hat{\sigma}^z_j + \frac{\tilde{U}}{4}\sum_{j} (\hat{\sigma}^z_j \hat{\sigma}^z_{j+1} + \frac{1}{2}\hat{\sigma}^z_j \hat{\sigma}^z_{j+2}) \notag \\
&-& \frac{U_2}{\sqrt{2}}\sum_{j} \hat{P}_{j-1} \hat{\sigma}^x_j \hat{P}_{j+1}, \label{Hs}
\end{eqnarray}
with $\hat{P}_j = (1-\hat{\sigma}^{z}_j)/2$ the projection operator which projects out the cases of two neighboring spins being polarized up simultaneously. This projection is necessary to make sure that the system remains in the $G_j=0$ sector and the resulting states can be described by the three allowed configurations shown in Fig.~\ref{Fig2}(a). Particularly at $\tilde{U}=m=0$ (denoted by the diamond in Fig.~\ref{Fig2}(b)), $\hat{H}_s$ reproduces the PXP model \cite{Turner2018A} which was originally realized in a Rydberg chain \cite{Bernien2017}. The PXP Hamiltonian carries a symmetry $\hat{\chi} \hat{H}_s \hat{\chi} = -\hat{H}_s$ with $\hat{\chi} = \prod_i \hat{\sigma}_j^z$. As a result, the energy spectrum is symmetric about $\epsilon=0$. This symmetry corresponds to the chiral symmetry of the original lattice gauge model \cite{SM}. The PXP model is well known to lead to dynamical revivals which refers to the phenomenon that the post-quench evolutions of the Rydberg $|\mathbb{Z}_2\rangle$ and $|\mathbb{Z}_3\rangle$ charge density waves (CDW) exhibit periodic recoveries and slow thermalization \cite{Bernien2017}.  This revival can be attributed to the quantum many-body scar states \cite{Turner2018A,Turner2018B,Serbyn2021}, which are the low-entropy eigenstates of the PXP Hamiltonian that violate the eigenstate thermalization hypothesis.

The $|\mathbb{Z}_2\rangle$ and $|\mathbb{Z}_3\rangle$ ordered states in our current model correspond exactly to the Rydberg $|\mathbb{Z}_2\rangle$ and $|\mathbb{Z}_3\rangle$ CDW states, and hence our model would also exhibit the dynamical revivals by quenching these two states into the chiral point. We perform such numerics by exactly diagonalizing (ED) the effective Hamiltonian $\hat{H}_\text{eff}$, and plot the evolution of the Loschmidt echo $\mathcal{L}(t)=\left\vert \left\langle \psi(0)|\psi (t)\right\rangle\right\vert ^{2}$ and the occupation on one gauge site in Fig.~\ref{Fig3} (a1) and (b1), where $|\psi(0)\rangle$ is initialized by $|\mathbb{Z}_2\rangle$ (solid line) and $|\mathbb{Z}_3\rangle$ (dashed line) states, respectively. The periodic oscillating curves clearly demonstrate the dynamical revivals. In Fig.~\ref{Fig3} (c1), we show the eigen-spectrum $\epsilon_i$ with vertical axis $\mathcal{S}$ denoting the bipartite entropy of eigenstates $|\epsilon_i\rangle$, where the scar states responsible for the $|\mathbb{Z}_2\rangle$- and $|\mathbb{Z}_3\rangle$-revivals are marked by cross and plus signs, respectively. These scars are selected according to the projective probability $|\langle \epsilon_i |\mathbb{Z}_2\rangle|^2$ and $|\langle \epsilon_i |\mathbb{Z}_3\rangle|^2$ above the threshold $\ge0.03$. As one can see, the scar states possess equal energy intervals and relatively low entropy within the spectrum. The energy interval $\Delta\epsilon$ matches well with the revival period $T$ of $\mathcal{L}(t)$ via the relation $T=2\pi/\Delta\epsilon$. In comparison, the dynamics of the same quantities, quenched to the Potts critical point, are plotted in Fig.~\ref{Fig3} (a2) and (b2), where the physical quantities exhibit fast thermalization without oscillation. To check the validity of these results, we also carry out numerical calculations based on the original Hamiltonian Eqs.~(\ref{Hint}) and (\ref{H0R}) using the technique of matrix product states \cite{Scholl2011}. The results are in excellent agreement with the ED results when the condition $\lambda_0\simeq q \gg U_0$ is satisfied \cite{SM}. This also serves as a confirmation of the validity of the effective Hamiltonian $\hat{H}_\text{eff}$.

The origin of the scars is of tremendous interest.
Recently, Yao and co-workers observed that the $|\mathbb{Z}_2\rangle$-related quantum scars migrate from the low-energy low-entropy states of the Ising transition \cite{Yao2021}. Considering the diagram Fig.~\ref{Fig2} (b) possesses a Potts criticality, hence now we have an opportunity in tracing the origin of the scar states associated with the $|\mathbb{Z}_3\rangle$ dynamics. We focus on the line $\tilde{U}=-2m$ (see the dot-dashed line in diagram Fig.~\ref{Fig2}(b)) and show the $|\mathbb{Z}_2\rangle$- as well as the $|\mathbb{Z}_3\rangle$-related scar spectra at the chiral-symmetric (diamond), middle (triangle) and Potts critical (star) points in Figs.~\ref{Fig3} (c1)-(c3), respectively. One may immediately observe that the spectra (c2) and (c3) are asymmetric about $\epsilon=0$ due to the chiral symmetry breaking induced by the finite $\tilde{U}$ interaction. Furthermore, as the Potts critical point is approached, $|\mathbb{Z}_3\rangle$- and $|\mathbb{Z}_2\rangle$-related scars respectively transfer to the low- and high-energy regimes, indicating that the scars associated with $|\mathbb{Z}_3\rangle$ originate from the low-energy low-entropy states of the Potts transition.

\textit{Summary and discussion---}
We proposed a scheme to synthesize the U(1) gauge invariance in a spin-1 Bose gas. The effective model exhibits a matter-field interaction which gives rise to a new $\mathbb{Z}_3$-ordered phase. This ordered phase connects to the disordered phase by the Potts criticality whose low-energy eigenstates are found to be the origin of quantum scar states responsible for the anomalous dynamical revivals of the $|\mathbb{Z}_3\rangle$ states. However, several questions remain unclear at the current stage. For example, what is the interpretation of the emerged matter-field interaction in particle physics? Why do the ordered states, $|\mathbb{Z}_2\rangle$ and $|\mathbb{Z}_3\rangle$, tend to be thermalized at quantum criticality? These questions will be addressed in the near future. Two very recent works \cite{Halimeh2022,Cheng2022} have explored the possibility of tuning the topological angle in the lattice Schwinger model. It will be also interesting to consider the similar possibility in our model and study the combined effect of topological angle and the matter-field interaction.
\begin{acknowledgments}
L. C. would like to thank Shang Liu, Yanting Cheng, Xin Chen, and Zhiyuan Yao for insightful discussion. L. C. acknowledges supports from the NSF of China (Grants Nos. 12174236 and 12147215); H. P. acknowledges supports from the US NSF and the Welch Foundation (Grant No. C-1669).

\end{acknowledgments}


\begin{thebibliography}{99}
\bibitem{QLM} There are two states for a local gauge field, i.e., $\left|0\right\rangle$ and $\left|2\right\rangle$, which allows us to replace the creation/annihilation ($\hat{b}^\dagger_{j,0} \hat{b}^\dagger_{j,0}$ and $\hat{b}_{j,0} \hat{b}_{j,0}$) by the spin-1/2 raising/lowering operators $\sqrt{2} \sigma^\pm_j$ with the factor $\sqrt{2}$ accounting for the Bose enhancement. Note that, in some literatures, the lattice gauge models with gauge bosons embedded in finite-dimensional Hilbert space are also called the quantum link models \cite{QLM2}.
\end{thebibliography}
\end{document}

\title{Supplemental Materials: Synthetic U(1) Gauge Invariance in a Spin-1 Bose Gas}
\author{Chunping Gao$^{1}$}
\author{Jinghu Liu$^{1}$}
\author{Maolin Chang$^{1}$}
\author{Han Pu$^{2}$}
\email{hpu@rice.edu}
\author{Li Chen$^{1}$}
\email{lchen@sxu.edu.cn}

\affiliation{
$^1${Institute of Theoretical Physics and State Key Laboratory of Quantum Optics and Quantum Optics Devices, Shanxi University, Taiyuan 030006, China}\\
$^2${Department of Physics and Astronomy, and Rice Center for Quantum
    Materials, Rice University, Houston, TX 77005, USA}
}

\maketitle

In this Supplemental Materials (SM), we provide additional information on this work. The SM is arranged as follows. In Sec.~\ref{SecI}, we show in detail the derivation of the effective Hamiltonian. In Sec.~\ref{SecII}, we discuss the relationship between our lattice model and the U(1) lattice Schwinger model, as well as how to transform our lattice model into a spin-1/2 chain. In Sec.~\ref{SecIII}, we discuss more on quantum phase transitions in the equilibrium phase diagram. In Sec.~\ref{SecIV}, quench dynamics based on the time-evolution of matrix product states (tMPS) is included.

\section{Effective Model} \label{SecI}
\begin{figure}[t]
\includegraphics[width=0.9\textwidth]{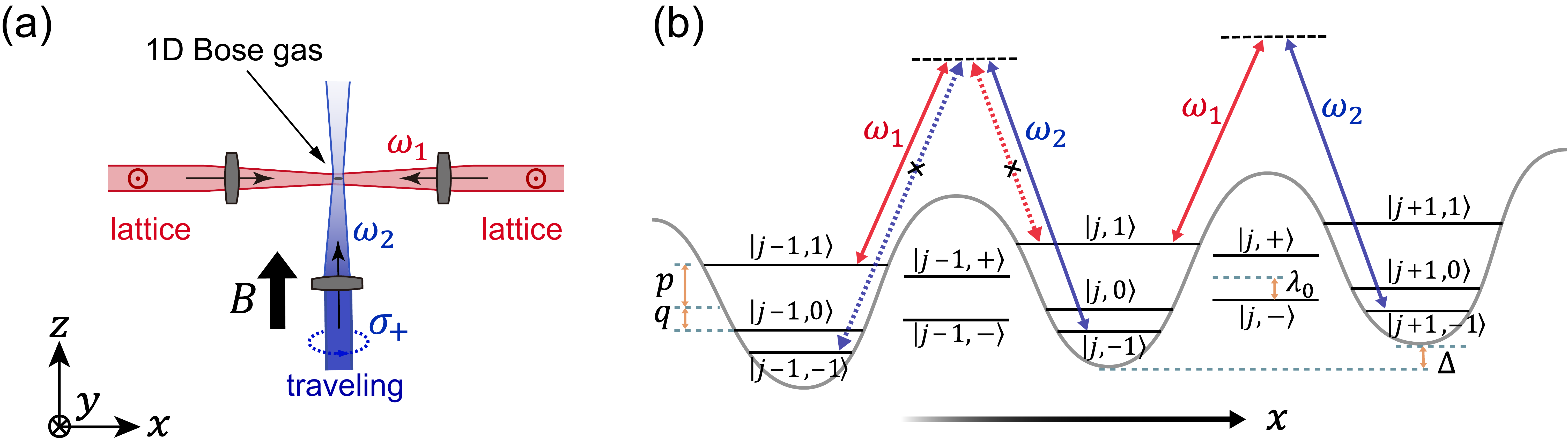}
\caption{(a) Detailed setup of our system, where the lattice beams (linear polarization along y) and the traveling light ($\sigma_+$ polarization) propagate along the $x$ and the $z$ directions, respectively. The quantized axis is fixed along the $z$ direction. (b) Atomic level structures. }
\label{FigS1}
\end{figure}

Our model system concerns a spin-1 Bose-Einstein condensate (BEC) confined in a one-dimensional optical lattice along direction $x$. In Fig.~\ref{FigS1} (a), we re-plot the architectures of our model with more details, which serves as a complement to Fig.~1(a) of the main text. The biased magnetic field $B$ fixes the quantization axis long the $z$-direction. The lattice beam propagates with frequency $\omega_1$ along the $x$-direction and is linearly polarized in the $y$-direction. To realize the laser-assisted hopping, we adopt another $\sigma_+$-polarized traveling wave along the $z$-direction with frequency $\omega_2$, which together with the lattice beam forms the $\Lambda$-type Raman transitions \cite{Miyake2013,liuxiongjun,panjianwei-2DSOC}, as schematically illustrated in Fig.~\ref{FigS1} (b). The Raman beams couple the states $\left\vert \pm 1\right\rangle$ while leaving the state $\left\vert 0 \right\rangle$ intact due to selection rule. Now, we have the single-particle Hamiltonian in the laboratory frame
\begin{equation}
\begin{split}
\hat{H}_{0}&=\frac{p_{x}^{2}}{2m}+V_{0}\sin ^{2}(k_{0}x) +\frac{\Delta }{a}x + M_{0}\sin (k_{0}x)(e^{-i\delta\omega t} |1\rangle \langle -1| + \text {h.c.}) \\
&\ \ \ +p\left(|1\rangle \langle 1|-|-1\rangle \langle -1|\right)+q\left(|1\rangle \langle 1|+|-1\rangle \langle -1|\right),
\end{split}
\label{initial}
\tag{S1}
\end{equation}%
where the first two terms represent the kinetic energy and the lattice potential of the neural atoms, respectively, with $k_0=\omega_1/c$ the wave vector of the lattice beam. $\Delta$ is a tilt field with $a=k_0/\pi$ the lattice constant, which can be realized using either the gravity \cite{gravity-tilt} or another optical trap with intensity gradient \cite{intensity-gradient}. The 4th term represents the Raman-assisted hopping, where $M_0$ is the two-photon Rabi frequency. The final two terms are the linear and the quadratic Zeeman splittings resulting from the biased field $B$.

Under the tight-binding approximation, we expand the above Hamiltonian in the lowest-band Wannier basis $|j,\sigma \rangle$, where $j$ and $\sigma$ indicate the site and the spin degrees of freedoms, respectively. Then we obtain
\begin{equation}
\begin{split}
\hat{H}_{0} &=-t\sum_{j,\sigma }\left( |j,\sigma \rangle \langle j+1,\sigma
|+ \text {h.c.}\right) \\
&\ \ \ +p\sum_{j}\left( |j,1\rangle \langle
j,1|-|j,-1\rangle \langle j,-1|\right)+q\sum_{j}\left( |j,1\rangle \langle j,1|+|j,-1\rangle \langle
j,-1|\right)  \\
&\ \ \ +\sum_{j,\sigma}j\Delta |j,\sigma\rangle \langle j,\sigma|+\sum_{j}\lambda_j \left( e^{-i\delta\omega t} |j,1\rangle \langle j+1,-1| + \text {h.c.}\right),
\end{split}
\tag{S2}
\end{equation}%
where the natural (spin-independent) hopping has been ignored due to the deep lattice confinement. $\lambda_j = (-1)^j\lambda_0$ characterizes the amplitude of the nearest-neighboring Raman-assisted hopping, which couples the states $|j,\sigma = 1\rangle$ and  $|j+1,\sigma = -1\rangle$, with $\lambda_0 =M_{0}\int dxw^{\ast }(x)\sin (k_{0}x)w(x-a)$ the overlap integral and $w(x)$ the Wannier function in coordinate representation. The staggered sign $(-1)^j$ can be attributed to the $2a$ period of the Raman coupling (see Eq.~(\ref{initial})), and hence $M_0\sin(k_0x)$ is antisymmetric relative to the minima of the local potential well \cite{liuxiongjun}. The off-resonant Raman coupling between the states $|j,1\rangle$ and $|j-1,-1\rangle$ has been neglected, as indicated in Fig.~\ref{FigS1}(b). Additionally, the Raman configuration would not result in on-site spin flips since the on-site integral naturally vanishes, i.e., $\int dxw^{\ast }(x)\sin (k_{0}x)w(x)=0$.

To discuss the effect of atomic interactions, we use the second quantization  formalism. The second-quantized $\hat{H}_{0}$ is given in Eq.~(2) of the main text. Under the unitary transformation
\begin{equation}
\hat{U}(t) =\prod_j \exp \left[i\left(j \Delta t\right)\hat{n}_j +i\left(\frac{\delta\omega t}{2}+\frac{\Delta t}{2}\right)\hat{F}^z_j\right],
\label{Ut}
\tag{S3}
\end{equation}
 we enter the rotating frame. Clearly, $\hat{U}(t)$ contains two parts of contribution: the former related to $\hat{n}_j$ is a spin-independent rotation which can eliminate the gradient term, while the latter with respect to $\hat{F}^z_j$ is to gauge out the off-diagonal phase factor $e^{i\delta\omega t}$ in $\hat H_0$ and the remnant phase factor $e^{i\Delta t}$ due to former rotation.
Moreover, another transformation $\hat{b}_{j,1}\rightarrow e^{i(j\pi) }\hat{b}_{j,1}$ and $\hat{b}_{j,0}\rightarrow e^{i(j\pi /2)}\hat{b}_{j,0}$ helps to eliminate the staggered sign in $\lambda_j$ and rewrite the single-particle Hamiltonian into
\begin{equation}
\begin{aligned}
\hat{H}_{0}&= p'\sum_{j}\hat{F}_j^z+q\sum_{j}\hat{\Xi}_j+
\sum_{j}\lambda_0\left( \hat{b}_{j,1}^{\dagger }\hat{b}%
_{j+1,-1}+\text {h.c.}\right),
\end{aligned}
\label{nth0}
\tag{S4}
\end{equation}%
which is Eq.~(4) of the main text. At $p'=0$, or namely $\delta\omega=2p-\Delta$, the Raman resonance occurs. Thus far, the total Hamiltonian is given by $\hat{H} = \hat{H}_0 + \hat{H}_\text{int}$, with \begin{equation}
\hat{H}_\text{int}=\frac{U_{0}}{2}\sum_{j}\hat{n}_{j}\left( \hat{n}_{j}-1\right) +\frac{U_{2}}{2}\sum_{j}\left( \hat{\mathbf{F}}_{j}^{2}-2\hat{n}_{j}\right)
\label{Hint}
\tag{S5}
\end{equation}
being the spin-1 interaction Hamiltonian also shown in Eq.~(3) of the main text.

The resonant Raman coupling opens up an energy gap $2\lambda_0$ for the two dressed modes $|j,\pm\rangle$ (associated with annihilation operators $\hat{a}_{j,\pm}$, respectively). These two dressed modes are related to the bare spin modes by Eq.~(5) of the main text. We remark that the lower-lying dressed modes $|j,-\rangle$ are treated as the matter fields in our lattice gauge model, and the spin-0 modes $|j,0\rangle$ the gauge fields. In the parameter regime $\lambda_0\simeq q \gg U_0$ [namely the restriction i) indicated in the main text], only the matter fields, $|j-1,-\rangle$ and $|j,-\rangle$, are resonantly coupled to the gauge field $|j,0\rangle$ satisfying $E_{j-1,-}+E_{j,-}\simeq 2 E_{j,0}$; whereas all the other processes involving the upper mode $|j,+\rangle$ are far-off resonant, as clearly illustrated in Fig.~\ref{FigS1}(b). This condition suppresses the unwanted two-body scatterings (scattering out of the matter and gauge modes). Hence in our calculation, we expand the total Hamiltonian $\hat{H}$ by the dressed modes $\hat{a}_{j,\pm}$, exclude all the $\hat{a}_{j,+}$-related interactions, and then obtain
\begin{equation}
\begin{aligned}
\hat{H}=& \left(q-\lambda_0\right)\sum_{j}\hat{N}_{j} + \frac{\tilde{U}}{2}\sum_{j}\hat{N}%
_{j-1}\hat{N}%
_{j}-\frac{U_{2}}{2}\sum_{j}\left(\hat{a}_{j-1,-}\hat{b}_{j,0}^{\dagger }\hat{b}_{j,0}^{\dagger }\hat{a}_{j,-}+\text{h.c.}\right) \\
&+\frac{U_{0}+U_{2}}{4}\sum_{j} \hat{N}_{j}(\hat{N}_{j}-1)+\frac{U_{0}}{2}\sum_{j} \hat{n}_{j,0}(\hat{n}_{j,0}-1)
\end{aligned}
\label{H2}
\tag{S6}
\end{equation}
with $\tilde{U}=(U_0-U_2)/2$ and $\hat{N}_{j} = \hat{a}_{j,-}^\dagger \hat{a}_{j,-}$, from which one can already observe that $\hat{H}$ is locally symmetric with respect to the Gauss operator $\hat G_j=\hat{N}_{j}+(\hat{n}_{j+1,0}+\hat{n}_{j,0})/2-1$. Here, we emphasize that the U(1) gauge symmetry is protected by the energy gap $2\lambda_0$ (or namely the restriction i)), without which the conservation of $\hat G_j$ will be broken. In Sec.~\ref{SecIV} of this SM, we perform a fully numerical calculation of the quench dynamics using the original Hamiltonian Eqs.~(\ref{nth0}) and (\ref{Hint}) with all the modes in presence. There, the effects of the restriction i) on $\hat G_j$ are demonstrated.

To restrict our system within the gauge sector $G_j=0$, we adopt the restriction ii) as noted in the main text, which helps to further simplify the last two terms of Eq.~(\ref{H2}) in the following way: the former term vanishes if the matter fields are limited to at most single occupation at each site; the latter one can be recast to $(U_{0}/2)\sum_{j}\hat{n}_{j,0}$ and merge with the single-particle term $(q-\lambda_0)\sum_{j}\hat{N}_{j} $ as the gauge fields are restricted within the empty or the doubly occupied subspace, i.e., $(\hat n_{j,0}-1)^2=1$. Eventually, the effective Hamiltonian $\hat{H}_\text{eff}$ [Eq.~(6) of the main text] can be obtained. From experimental point of view, various developed techniques can be used for the manipulation of particle numbers. For example, in order to prepare the $|\mathbb{Z}_2 \rangle = |\cdots 2_0 0_0 2_0 \cdots\rangle$ state, one may first prepare the system in the Mott state of spin-0 component with $ n_{j,0}=2$ and then blow the particles on even sites away using site-selective addressing approach \cite{panjianwei-siteselect}.

\section{Model Correspondence} \label{SecII}
\subsection{Correspondence to the Lattice Schwinger Model}
The (1+1)-dimensional lattice Schwinger model (LSM) is a paradigm of simulating the U(1) lattice gauge theory in QED, whose Hamiltonian is given by \cite{Zoller2012}
\begin{equation}
\hat{H}_\text{LSM} = m\sum_j (-1)^j \hat{\psi}^\dagger_j \hat{\psi}_j + \frac{g^2}{2} \sum_j \hat{E}_{j,j+1}^2 -t \sum_j \hat{\psi}^\dagger_j \hat{U}_{j,j+1} \hat{\psi}_{j+1} ,
\label{HLSM}
\tag{S7}
\end{equation}
where $\hat{\psi}_j$ is the fermionic matter field on site $j$, $\hat{U}_{j,j+1}$ and $\hat{E}_{j,j+1}$ are the bosonic gauge field on the link $(j,j+1)$ which satisfies $[\hat{E}_{j,j+1},\hat{U}_{k,k+1}] = \delta_{jk}\hat{U}_{k,k+1}$ \cite{Surace2020}. $m$, $g^2$ and $t$ represent, respectively, the staggered mass of fermions, the amplitude of gauge coupling and the intensity of matter-gauge interaction. To approach the continuum limit of the field theory, the Hilbert space of the gauge field should go to infinity \cite{Coleman1976}. However, in practical quantum simulations of the LSM \cite{Martinez2016}, the gauge field has to be limited in a finite-dimentional Hilbert space, which is also called the quantum link model \cite{QLM}. In the following, we shall demonstrate that our effective Hamiltonian $\hat{H}_\text{eff}$ [Eq.~(6) of the main text] is equivalent to the LSM with gauge fields realized by Pauli spins.

First, we apply the Jordan-Wigner transformation to transform the bosonic matter fields to fermions, i.e.,
\begin{equation}
\begin{aligned}
\hat \psi_j&=\hat a_{j,-}\prod_{l=1}^{j-1}(2\hat N_l-1),\\
\hat \psi_j^{\dag}&=\hat a_{j,-}^{\dagger}\prod_{l=1}^{j-1}(2\hat N_l-1)
\end{aligned}
\tag{S8}
\end{equation}
with $\hat N_l=\hat a_{l,-}^{\dagger}\hat a_{l,-}$, which leads to
\begin{equation}
\hat{H}=m\sum_{j}
\hat \psi_j^\dag\hat \psi_j+\frac{\tilde{U} }{2}\sum_{j}\hat \psi_{j-1}^\dag\hat \psi_{j-1}\hat \psi_j^\dag\hat \psi_j+\frac{U_{2}}{2}\sum_{j}\left(\hat \psi_j \hat{b}_{j,0}^{\dagger }\hat{b}_{j,0}^{\dagger
} \hat \psi_{j-1}+\text{h.c.}\right).
\tag{S9}
\label{Hf1}
\end{equation}
Next, we rewrite the gauge fields using spin-1/2 Pauli spins in the manner of
\begin{equation}
\begin{aligned}
\hat\sigma _{j}^{+}&\leftrightarrow \frac{1}{\sqrt{2}}\hat{b}_{j,0}^{\dagger }\hat{b}_{j,0}^{\dagger
},\\ \hat\sigma_{j}^{-}&\leftrightarrow\frac{1}{\sqrt{2}}\hat{b}_{j,0}\hat{b}_{j,0}, \\
\hat\sigma_{j}^{z}&\leftrightarrow\hat{n}_{j,0}-1,
\end{aligned}
\label{spin}
\tag{S10}
\end{equation}
which recasts Eq.~(\ref{Hf1}) into
\begin{equation}
\hat{H}=m\sum_{j}
\hat \psi_j^\dag\hat \psi_j+\frac{\tilde{U} }{2}\sum_{j}\hat \psi_{j-1}^\dag\hat \psi_{j-1}\hat \psi_j^\dag\hat \psi_j+\frac{U_{2}}{\sqrt{2}}\sum_{j}\left(\hat \psi_j\hat \sigma _{j}^{+}\hat \psi_{j-1}+\text{h.c.}\right).
\tag{S11}
\end{equation}
A subsequent particle-hole transformation on the odd-site matter fields $\hat \psi_{j\in\text{odd}}\rightarrow\hat \psi_{j\in\text{odd}}^\dagger$ and the corresponding gauge flips on the odd sites $\hat \sigma_{j\in\text{odd}}^+\rightarrow-\hat \sigma_{j\in\text{odd}}^-,\hat \sigma_{j\in\text{odd}}^z\rightarrow-\hat \sigma_{j\in\text{odd}}^z$ eventually lead to
\begin{equation}
\begin{aligned}
\hat{H}_\text{f} &= \frac{\tilde{U}}{2}\sum_{j}%
\hat \psi_j^\dag\hat \psi_j+\left(\frac{\tilde{U}}{2}+m\right) \sum_{j} (-1)^j  \hat{\psi}_j^\dagger \hat{\psi}_j\\
&\ \ \ -\frac{\tilde{U}}{2}\sum_{j} \hat{\psi}_{j-1}^\dagger \hat{\psi}_{j-1} \hat{\psi}_{j}^\dagger \hat{\psi}_{j}
-\frac{U_2}{\sqrt{2}}\sum_{j} \left(\hat{\psi}_{j-1}^\dagger \hat\sigma_j^+ \hat{\psi}_{j}+ \text{h.c.}\right).
\end{aligned}
\label{Hf0}
\tag{S12}
\end{equation}
Following the same manner, we obtain the corresponding Gauss's Law in terms of the fermions and the Pauli spins, i.e.,
\begin{equation}
\hat G_j=\hat \psi_j^\dag\hat \psi_j - \frac{\hat\sigma_{j+1}^z-\hat\sigma_{j}^z}{2} + \frac{1}{2}[(-1)^j-1].
\label{Gf}
\tag{S13}
\end{equation}
One may immediately notice that, except for the first term, Eq.~(\ref{Hf0}) reproduces the Eq.~(8) of the main text. In a fixed gauge sector, the first term is simply a constant that can be safely dropped out using the Gauss's Law Eq.~(\ref{Gf}).

In the case of $\tilde{U}=0$, $\hat{H}_\text{f}$ in Eq.~(\ref{Hf0}) exactly recovers the LSM Eq.~(\ref{HLSM}) with $t\leftrightarrow U_2/\sqrt{2}  $ and $\hat{E}^2_{j,j+1}\leftrightarrow(\hat\sigma^z_{j})^2=1$. Particularly for the case of $\tilde{U}=m=0$, $\hat{H}_\text{f}$ exhibits a chiral symmetry $\mathcal{P}\mathcal{T}(\hat{H}_\text{f})=-\hat{H}_\text{f}$ with $\mathcal{P}$ and $\mathcal{T}$ denoting the particle-hole transformation and one-site translation, respectively. As a consequence, the spectra are symmetric about $\epsilon=0$. The chiral symmetry would be broken by either the staggered mass $m$ or the fermionic interaction $\tilde{U}$.

\subsection{Correspondence to the spin-1/2 chain}
The Gauss's Law connects matter and gauge fields as the gauge is fixed, implying that one of them is a redundant degree of freedom.
In this work, we concentrate on the gauge sector $G_j=0$, which allows us to integrate the matter fields out and get an effective spin-1/2 chain model.

We start the calculation with the effective Hamiltonian $\hat{H}_\text{eff}$ Eq.~(6) and the Gauss's Law $\hat G_j$ Eq.~(7) of the main text. Substituting Eq.~(\ref{spin}) into $\hat G_j$, the Gauss operator can be expressed as
\begin{equation}
\hat G_j=\hat{N}_{j}+\frac{\hat{\sigma}_{j+1}^z+\hat{\sigma}_{j}^z}{2}.
\label{G}
\tag{S14}
\end{equation}
In the gauge sector $G_j=0$, the effective Hamiltonian $\hat{H}_\text{eff}$ has the following operator correspondence \cite{Yao2021}
\begin{equation}
\begin{aligned}
\hat{N}_j \leftrightarrow -(\hat{\sigma}_j^z+\hat{\sigma}_{j+1}^z)/2, \\
\hat{a}_{j-1,-}\hat{\sigma}_j^+\hat{a}_{j,-} +\hat{a}_{j,-}^{\dag}\hat{\sigma}_j^-\hat{a}_{j-1,-}^{\dag} \leftrightarrow \hat{\sigma}_j^x,
\end{aligned}
\label{map}
\tag{S15}
\end{equation}
under which we obtain the mapped spin model
\begin{equation}
\hat{H}_\text{s} = -m \sum_{j} \hat{\sigma}^z_j + \frac{\tilde{U}}{4}\sum_{j} (\hat{\sigma}^z_j \hat{\sigma}^z_{j+1} + \frac{1}{2}\hat{\sigma}^z_j \hat{\sigma}^z_{j+2})- \frac{U_2}{\sqrt{2}}\sum_{j} \hat{\sigma}^x_j.
\label{Hs}
\tag{S16}
\end{equation}
Notably, the first line of the mapping Eq.~(\ref{map}) rigorously comes from the Gauss's Law, while the second line simply drops the matter fields out. To guarantee the equivalence of the mapping, an additional constraint should be added: two nearby gauge fields cannot be polarized up simultaneously.
This constraint is equivalent to projecting out the configurations not within the $G_j=0$ sector during spin flips, i.e., $\hat{\sigma}^x_j \rightarrow \hat{P}_{j-1} \hat{\sigma}^x_j \hat{P}_{j+1}$ where $\hat{P}_j = (1-\hat{\sigma}^{z}_j)/2$ denotes the projection operator. As a result, the main text's Eq.~(10) can be achieved. Particularly in the chiral-symmetric point $\tilde{U}=m=0$, $\hat{H}_s$ is known as the PXP Hamiltonian which was firstly observed in a Rydberg quantum simulator \cite{Bernien2017}. The PXP Hamiltonian has attracted tremendous attentions in the study of slow dynamics, many-body revivals, and quantum many-body scars \cite{Turner2018A,Turner2018B}.

\section{Phase transitions in equilibrium} \label{SecIII}
\begin{figure}[t]
\includegraphics[width=0.88\textwidth]{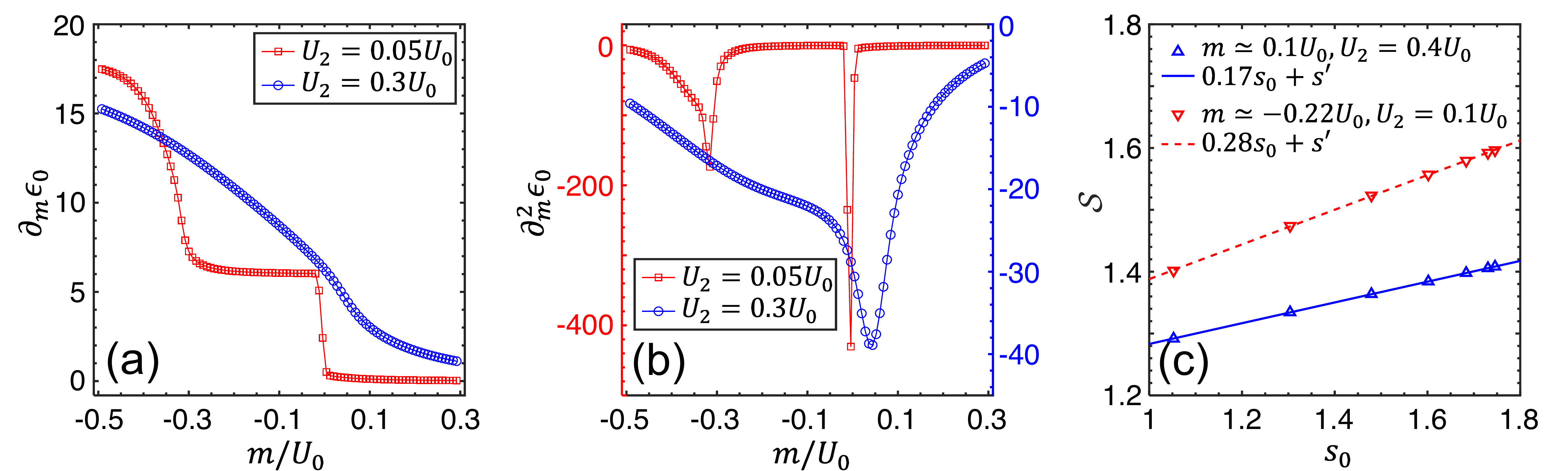}
\caption{ (a) First-order derivatives of the ground-state energy $\partial_{m} \epsilon_0$, where we fix $U_2$ and sweep $m$. (b) Second-order derivatives of the ground-state energy $\partial^2_{m} \epsilon_0$. (c) Von Neumann entropy versus the different subsystem size, respectively at the D-$\mathbb{Z}_2$ critical point ($m\simeq0.1U_0,U_2=0.4U_0$) and at the D-$\mathbb{Z}_3$ critical point ($m\simeq-0.22U_0,U_2=0.1U_0$). The line slopes relate to the central charge $c$ in the way of $0.17=c/3$ and $0.28=c/3$. In our calculation, we take $L=18$. }
\label{FigS2}
\end{figure}

In the main text, we display a rich phase diagram where various phase transitions are identified. Here, we discuss the identification of phase transitions with further detail. According to the Ehrenfest classification \cite{Jaeger1998}, a n$^\text{th}$-order phase transition would exhibit a discontinuity of the n$^\text{th}$-order derivative of the free energy. At zero temperature, the free energy coincides with the internal energy which can be calculated by diagonalizing the Hamiltonian $\hat{H}_{\text{eff}}$. In Fig.~\ref{FigS2} (a) and (b), we respectively show the 1st-order and the 2nd-order derivatives of the ground-state energy as functions of $m$. In each subfigure, the line with squares denotes the case of $U_2=0.05U_0$, while the line with circles denotes the case of $U_2=0.3U_0$. The discontinuity of $\partial_m \epsilon_0$ in Fig.~\ref{FigS2} (a) clearly indicates the 1st-order $\mathbb{Z}_2$-$\mathbb{Z}_3$ transition, while the discontinuities of $\partial_m^2 \epsilon_0$ help to identify the 2nd-order D-$\mathbb{Z}_2$ and D-$\mathbb{Z}_3$ transitions, respectively.

On the 2nd-order critical points, we classify the transitions by calculating the central charge $c$, which can be realized through fitting the von Neumann entropy $\mathcal{S}$ (Eq.~(9) of the main text) as a function of subsystem size $l_A$. In Fig.~\ref{FigS2}(c), we show $\mathcal{S}$ at the D-$\mathbb{Z}_2$ critical point ($m\simeq0.1U_0,U_2=0.4U_0$) and the D-$\mathbb{Z}_3$ critical point ($m\simeq-0.22U_0,U_2=0.1U_0$), respectively. These two points are marked by the cross and the star on the diagram Fig.~2(b) in the main text. It is clearly shown that $\mathcal{S}$ behaves linearly with respect to $s_0=\ln\left[\frac{L}{\pi}\sin\left(\frac{\pi l_A}{L}\right)\right]$, and the line slopes yield the central charge $c/3$. Furthermore, to avoid the size effect of the system size $L$, we make the finite-scaling analysis with respect to $c$, which is shown in Fig.~2(c) of the main text, where the extrapolated $c$ can be obtained in the thermodynamic limit $1/L\rightarrow0$.

\section{Quench Dynamics using TMPS} \label{SecIV}
\begin{figure}[t]
\includegraphics[width=0.75\textwidth]{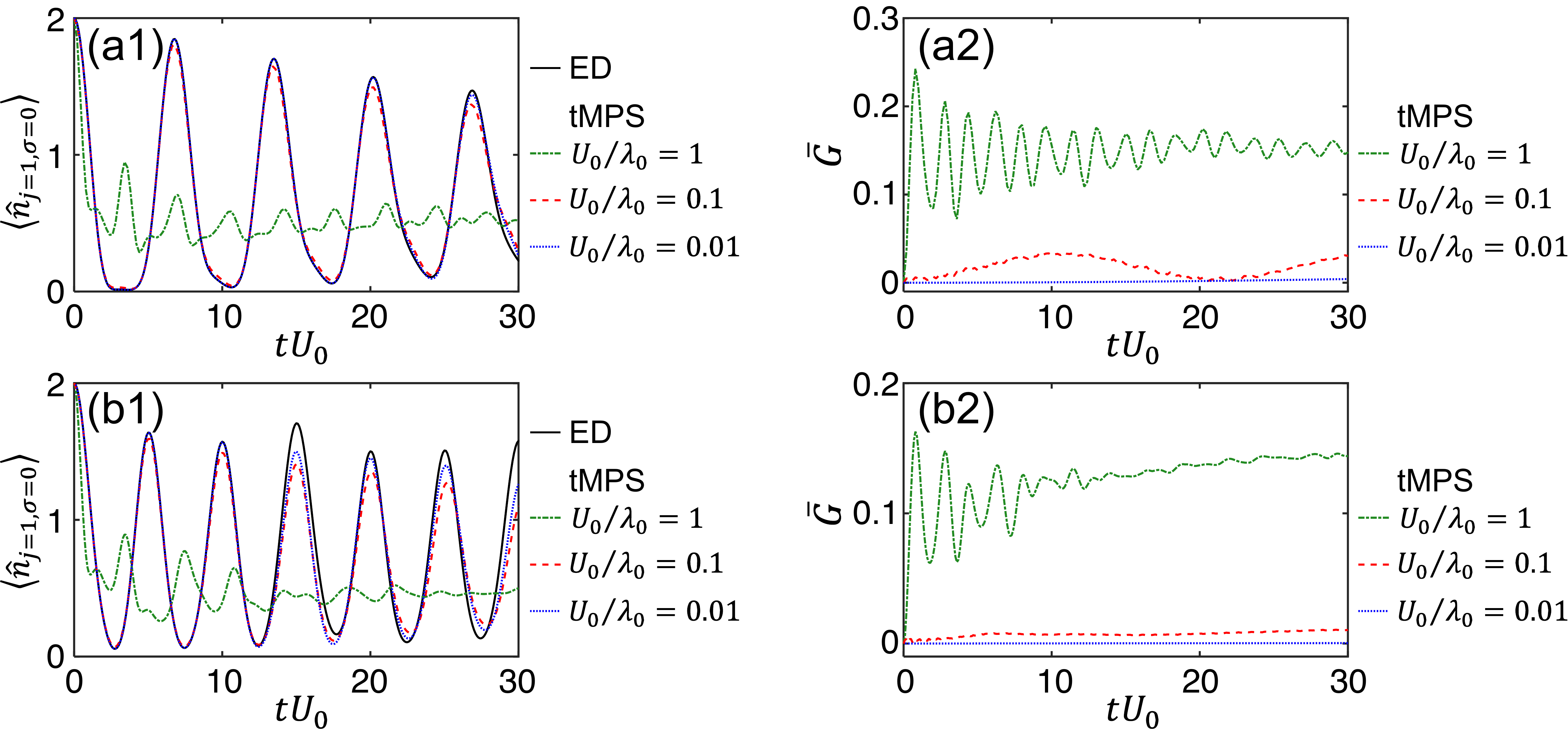}
\caption{$|\mathbb{Z}_2\rangle$-state (a) and $|\mathbb{Z}_3\rangle$-state (b) post-quench dynamics at the chiral-symmetric point $\tilde{U}=m=0$. (a1) and (b1) Evolution of the particle number of the first gauge site, i.e., $\langle \hat{n}_{j=1,\sigma=0}\rangle$. (a2) and (b2) Evolution of the averaged expectation of the Gauss operator $\bar{G}$. In each subfigure, the solid line denotes the ED's result based on $\hat{H}_\text{eff}$ obtained under the assumption ${\bar{G}}=0$, while the dot-dashed, dashed, and dotted lines, respectively corresponding to the cases $U_0/\lambda_0=1,0.1,0.01$, indicate the results obtained by the tMPS method using the original Hamiltonian Eqs.~(\ref{nth0}) and (\ref{Hint}). We fix $L=54$ and $L=18$ for the tMPS and ED calculations, respectively.}
\label{FigS3}
\end{figure}

As mentioned in Sec.~\ref{SecI}, the restriction i) [$\lambda_0\simeq q \gg U_0$] helps to eliminate the off-resonant $\hat{a}_{j,+}$ modes, which thus stands as the prerequisite for our effective Hamiltonian $\hat{H}_\text{eff}$ [Eq.~(6) of the main text] and the gauge invariance $\hat{G}_j$ [Eq.~(7) of the main text]. In this section, let us in detail discuss how this restriction would affect our lattice gauge model in the context of quench dynamics. To this end, we adopt the tMPS method \cite{Scholl2011} for the numerics based on the original Hamiltonian Eqs.~(\ref{nth0}) and (\ref{Hint}). In our calculation, we set the lattice length $L=54$ and quench the $|\mathbb{Z}_2\rangle$ and $|\mathbb{Z}_3\rangle$ states to the chiral point $\tilde{U}=m=0$. We focus on two operators that are experimentally measurable; one is the particle number of a single gauge site, e.g. $\langle \hat{n}_{j,\sigma=0}\rangle$, and the other is the averaged expectation of the Gauss operator, i.e.,
\begin{equation}
\bar{G} = \frac{1}{L}\sum_j \left| \left\langle\hat{G}_j \right\rangle\right|,
\label{AG}
\tag{S17}
\end{equation}
where $\langle\hat{G}_j\rangle$, according to the definition [Eq.~(7) of the main text], is simply the measurement on the local particle numbers within a building block.
In the gauge sector $G_j=0$, the deviation of $\bar{G}$ to zero reflects the breaking of the gauge invariance.
In Figs.~\ref{FigS3}(a) and (b), we show the dynamics of these two observables with respect to the $|\mathbb{Z}_2\rangle$- and $|\mathbb{Z}_3\rangle$-state quench dynamics. In each subfigure, the dot-dashed, dashed and dotted lines indicate the cases with $U_0/\lambda_0=1$, $U_0/\lambda_0=0.1$, and $U_0/\lambda_0=0.01$, corresponding to the situations that the restriction i) being unsatisfied, just satisfied, and strictly satisfied, respectively. In comparison, the dynamics of $\langle \hat{n}_{j,\sigma=0}\rangle$ obtained by exactly diagonalizing the effective Hamiltonian $\hat{H}_\text{eff}$ are also plotted by solid lines in Figs.~\ref{FigS3}(a1) and (b1). One can clearly observe that, for the two cases, $U_0/\lambda_0=0.1$ and $U_0/\lambda_0=0.01$ that the restriction i) is satisfied, the tMPS dynamics of $\langle \hat{n}_{1,0}\rangle$ are in good agreement with those obtained by the ED method, and $\bar{G}$ exhibits small oscillation around zero. In contrast, for the case of $U_0/\lambda_0=1$ where the restriction i) is broken, the resulting $\langle \hat{n}_{1,0}\rangle$ exhibits a large discrepancy to the ED's results, accompanied by the large deviation of the local conservation $\bar{G}$ from zero. These results serve as a criterion for the validity of our effective model $\hat{H}_\text{eff}$.